\documentclass[final,10pt]{scrartcl}
\pdfoutput=1

\usepackage[utf8]{inputenc}
\usepackage[T1]{fontenc}

\usepackage[verbose=true]{microtype}
\usepackage{tabularx}
\usepackage{booktabs}
\usepackage{color}
\usepackage[english]{babel}
\usepackage{hyperref}

\usepackage{slashed}
\usepackage{caption}
\usepackage{subcaption}
\usepackage{amsmath, amssymb}
\usepackage{amsthm}
\usepackage{xspace}
\usepackage{graphicx}

\newcommand{\rnum}[1]{\uppercase\expandafter{\romannumeral #1\relax}}

\newcommand{\fer}[1]{\psi_{#1}}

\newcommand{\gau}[1]{\lambda_{#1}}

\newcommand{\yuk}[2]{\ensuremath{\smash[t]{\Upsilon_{#1}^{\protect\phantom{#1}#2}}}\xspace}
\newcommand{\yuks}[2]{\ensuremath{\Upsilon_{#1}^{\phantom{#1}#2}}{}^*\xspace}

\newcommand{\yukw}[2]{\ensuremath{\widetilde{\Upsilon}_{#1}^{\protect\phantom{#1}#2}}\xspace}

\newcommand{\Cw}[1]{\widetilde C_{#1}}

\def\sfer{\widetilde{\psi}}

\newcommand{\BBBB}[1]{\ensuremath{\mathcal{B}_{\textrm{maj}}}\xspace} 
\newcommand{\BBB}[1]{\ensuremath{\mathcal{B}_{#1}}\xspace} 
\newcommand{\B}[1]{\ensuremath{\mathcal{B}_{#1}}\xspace}

\newcommand{\Bc}[2]{\ensuremath{\mathcal{B}_{#1}^{#2}}\xspace}

\def\bas#1\eas{\begin{align*}#1\end{align*}}
\def\ba#1\ea{\begin{align}#1\end{align}}
\def\bi#1\ei{\begin{itemize}#1\end{itemize}}
\def\be#1\ee{\begin{enumerate}#1\end{enumerate}}
\def\nn{\nonumber}

\newcommand{\bino}[1]{\ensuremath{\lambda_{0#1}}\xspace}
\newcommand{\wino}[1]{\ensuremath{\vec{\lambda}_{#1}}\xspace}
\newcommand{\gluino}[1]{\ensuremath{g_{#1}}\xspace}

\newcommand{\photon}[0]{\ensuremath{B}\xspace}
\newcommand{\weak}[0]{\ensuremath{W}\xspace}
\newcommand{\gluon}[0]{\ensuremath{g}\xspace}

\def\hu{\ensuremath{h_u}\xspace}
\def\hd{\ensuremath{h_d}\xspace}

\def\shu{\ensuremath{\widetilde{h}_u}\xspace}
\def\shd{\ensuremath{\widetilde{h}_d}\xspace}
\def\ashu{\ensuremath{\overline{\widetilde{h}}_u}\xspace}
\def\ashd{\ensuremath{\overline{\widetilde{h}}_d}\xspace}

\def\maj{\Upsilon_{\mathrm{m}}}

\def\A{\mathcal{A}}

\def\H{\mathcal{H}}
\def\K{\mathcal{K}}

\def\n{\mathcal{N}}

\def\com{\mathbb{C}}

\DeclareMathOperator{\diag}{diag}
\DeclareMathOperator{\tr}{tr}
\DeclareMathOperator{\id}{id}

\newcommand{\rep}[2]{\ensuremath{\mathbf{N}_{#1} \otimes \mathbf{N}_{#2}^{o}}\xspace}
\newcommand{\repl}[2]{\ensuremath{\mathbf{#1} \otimes \mathbf{#2}^o}\xspace}

\def\dirac{\ensuremath{\slashed{\partial}_M}\xspace}

\newcommand{\w}[1]{\ensuremath{\omega_{#1}}}

\def\sgnc{\epsilon}

\theoremstyle{plain}
\newtheorem{theorem}{Theorem}

\newtheorem{rmk}[theorem]{Remark}

\usepackage{import}

\newtoks\svgpath
\svgpath={gfx/}

\graphicspath{{gfx/}}

\newcommand{\includesvg}[1]{
\subimport{\the\svgpath}{#1.pdf_tex}
}

\newcommand{\fig}[6]{\begin{figure}[#1] \centering \def\svgwidth{#2} \subimport{\the\svgpath}{#3.pdf_tex}\captionsetup{width=#4}\caption{#5}\label{fig:#6} \end{figure}}

\usepackage{a4wide}
\usepackage[draft]{showkeys}

\usepackage{parskip}
\def\sfer{\ensuremath{\phi}\xspace}

\usepackage[affil-it]{authblk}

\author[a]{Wim Beenakker%
  \thanks{Electronic address: \texttt{W.Beenakker@science.ru.nl}}}
\author[a,b]{Thijs van den Broek%
  \thanks{Electronic address: \texttt{T.vandenBroek@science.ru.nl}}}
\author[a]{Walter D.~van Suijlekom%
  \thanks{Electronic address: \texttt{waltervs@math.ru.nl} (corresponding author)}}
\affil[a]{Radboud University Nijmegen, Institute for Mathematics, Astrophysics and Particle Physics,
Faculty of Science, PO Box 9010, 6500 GL, Nijmegen, The Netherlands}
\affil[b]{Nikhef, Science Park Amsterdam 105, 1098 XG Amsterdam}

\title{Supersymmetry and noncommutative geometry}
\subtitle{Part \rnum{3}: The noncommutative supersymmetric Standard Model}
\date{}

\begin{document}

\maketitle

\begin{abstract}
In a previous paper we developed a formalism to construct (potentially) supersymmetric theories in the context of noncommutative geometry. We apply this formalism to explore the existence of a noncommutative version of the minimal supersymmetric Standard Model (MSSM). We obtain the exact particle content of the MSSM and identify (in form) its interactions but conclude that their coefficients are such that the standard action functional used in noncommutative geometry is in fact not supersymmetric. 
\end{abstract}

\section{Introduction}

In \cite{BS13I} we provided a classification of potentially supersymmetric models within the framework of noncommutative geometry \cite{C94,CCM07}. The context in which this was performed were the almost-commutative geometries (ACG, \cite{ISS03}) of KO-dimension 2 on a flat, four-dimensional manifold. This classification entailed the identification of five building blocks: extensions of the ACG that have or retain a supersymmetric particle content and are necessary for keeping or making the action supersymmetric. In addition, we provided a list of sufficient demands for the standard action functional that is used in noncommutative geometry to actually be supersymmetric. Subsequently, the topic of soft supersymmetry breaking in this context (cardinal for constructing viable supersymmetric theories) was explored in \cite{BS13II}. \\

In this third paper of the series we focus on the minimal supersymmetric Standard Model (MSSM, see e.g. \cite{DGR04,CEKKLW05} for detailed accounts), phenomenologically the most important example of ($N= 1$) supersymmetry. This model encompasses all particles that the Standard Model features, but extended with their respective superpartners. In addition, demanding a theory that is free from anomalies and a superpotential that is holomorphic, it follows that we must distinguish between up-type and down-type Higgses (that get their name from whether they give mass to the up-type or down-type fermions only). The MSSM Higgses together thus have four complex degrees of freedom (which, after symmetry breaking, results in five real scalar particles, see \cite{DGR04}, Chapter 10). \\

We explore the possibilities for obtaining the particle content and action of the MSSM in this context. The paper is organised as follows. First we will provide a short recapitulation of the aforementioned classification in Section \ref{sec:recap}. In Section \ref{sec:NCSSM} we will list the basic properties of the almost-commutative geometry that is to give the MSSM, including the building blocks it consists of. To confirm that we are on the right track we identify all MSSM particles and examine their properties in Section \ref{sec:identification}. Finally, in Section \ref{sec:ncmssm-checks} we will confront our model with a sufficient number of the demands from \cite[\S 3]{BS13I} to verify that the action associated to this model is not supersymmetric. Throughout this paper, we will a priori allow for a number of generations other than $3$.

\section{Supersymmetry in noncommutative geometry}\label{sec:recap}

The context in which the classification of potentially supersymmetric theories was found, was a particular class of noncommutative geometries; the \emph{almost-commutative geometries} (\cite{S00}, see \cite{DS11} for an introduction),
\begin{align*}
	(C^{\infty}(M, \A_F), L^2(M, S \otimes \H_F), i\gamma^\mu \nabla^S_\mu + \gamma_M \otimes D_F, J_M \otimes J_F; \gamma_M \otimes \gamma_F).
\end{align*}
It is the tensor product of a (real, even) \emph{canonical spectral triple} \cite[Ch 6.1]{C94} with a (real, even) \emph{finite spectral triple}. With the first we mean the data
\begin{align*}
	(C^{\infty}(M), L^2(M, S), \dirac = i\gamma^\mu \nabla^S_\mu; J_M, \gamma_M),
\end{align*}
where $(M, g)$ is a compact Riemannian spin manifold, $L^2(M, S)$ denotes the square integrable sections of the corresponding spinor bundle and \dirac is the \emph{Dirac operator} that is derived from the Levi-Civita connection on $M$, $\gamma_M$ is the chirality operator (only for even-dimensional $M$) and $J_M$ denotes charge conjugation. Real, even finite spectral triples are all of the form 
\begin{align*}
	(\A_F, \H_F, D_F; J_F, \gamma_F)   
\end{align*}
where $\A_F$ is a (finite) direct sum of matrix algebras over $\mathbb{R}$, $\mathbb{C}$ or $\mathbb{H}$, $\H_F$ is a (finite dimensional) $\A_F$-bimodule, whose right module structure is implemented by a \emph{real structure} $J_F$ (i.e.~$ \xi a := J_Fa^*J^*_F\xi$, $\xi \in \H_F$, $a \in \A_F$), $\gamma_F$ is a grading (i.e.~$\gamma^2_F = 1, \gamma_F = \gamma^*_F$) and $D_F$ is a Hermitian matrix on $\H_F$. \\

There are several extra demands on the elements of spectral triples. These will not be covered here but can be found in e.g.~\cite[\S 1.1]{BS13I}. 
Finite spectral triples (and consequently almost-commutative geometries) can be classified using Krajewski diagrams \cite{KR97}. These are also covered in detail in \cite[\S 1.4]{BS13I}. \\ 


To each almost-commutative geometry we can associate a natural, gauge invariant action \cite{CC97}:
\begin{align}
	S[\psi, A] := \frac{1}{2}\langle J\psi, D_A \psi \rangle + \tr f(D_A/\Lambda),\qquad \psi \in \H^+,\label{eq:totalaction}
\end{align}
where $D_A$ is the total Dirac operator (i.e.~including its inner fluctuations \cite{C96}, \cite[\S \MakeUppercase{\romannumeral 11}]{C00}), $f$ must be a positive, even function, $\Lambda$ is an (a priori unknown) mass scale and with $\langle ., .\rangle$ we denote the inner product on $\H = L^2(M, S\otimes \H_F)$, whose input is restricted to spinors of $\gamma_M\otimes \gamma_F$--eigenvalue $+1$. This restriction is needed to avoid overcounting the fermionic degrees of freedom \cite{LMMS97, CCM07}, but requires $\gamma_M \otimes \gamma_F$ to anticommute with $J_M \otimes J_F$, i.e.~we require the KO-dimension of the full spectral triple to equal $2$ or $6$. We will restrict ourselves to four-dimensional manifolds and, following the success of the Standard Model from noncommutative geometry (NCSM), demand the finite spectral triple to have KO-dimension 6. The second term of \eqref{eq:totalaction}, called the \emph{spectral action}, is in the context of almost-commutative geometries typically handled by performing a heat kernel expansion \cite{Gil84} in $\Lambda$. For almost-commutative geometries on compact, flat, four-dimensional manifolds without boundary ---the objects we are studying here--- the first terms of this expansion read \cite{CC97, BS13I}:
\begin{align}
\tr f\bigg(\frac{D_A}{\Lambda}\bigg) &\sim \int_M \bigg[\frac{f(0)}{8\pi^2}\Big( - \frac{1}{3}\tr_F\mathbb{F}_{\mu\nu}\mathbb{F}^{\mu\nu} + \tr_F \Phi^4 + \tr_F [D_\mu, \Phi]^2\Big)\nonumber \\
	&\qquad + \frac{1}{2\pi^2}\Lambda^4 f_4\tr_F \id - \frac{1}{2\pi^2}\Lambda^2 f_2\tr_F\Phi^2\bigg]  + \mathcal{O}(\Lambda^{-2}), \label{eq:spectral_action_acg_flat}
\end{align}
where $f_{n}$ is the $(n-1)$th moment of the function $f$, $D_\mu := \partial_\mu + \mathbb{A}_\mu$ is the covariant derivative, with $\tr_F$ we mean the trace over the finite Hilbert space and $\mathbb{F}_{\mu\nu}$ denotes the (anti-Hermitian) field strength (or curvature) that corresponds to $\mathbb{A}_\mu$. Thus, physically, the Hilbert space $\H$ contains all fermionic data, the gauge bosons are generated by the canonical Dirac operator \dirac and the scalar fields (contained in $\Phi$) are generated by $D_F$. \\

\begin{figure}
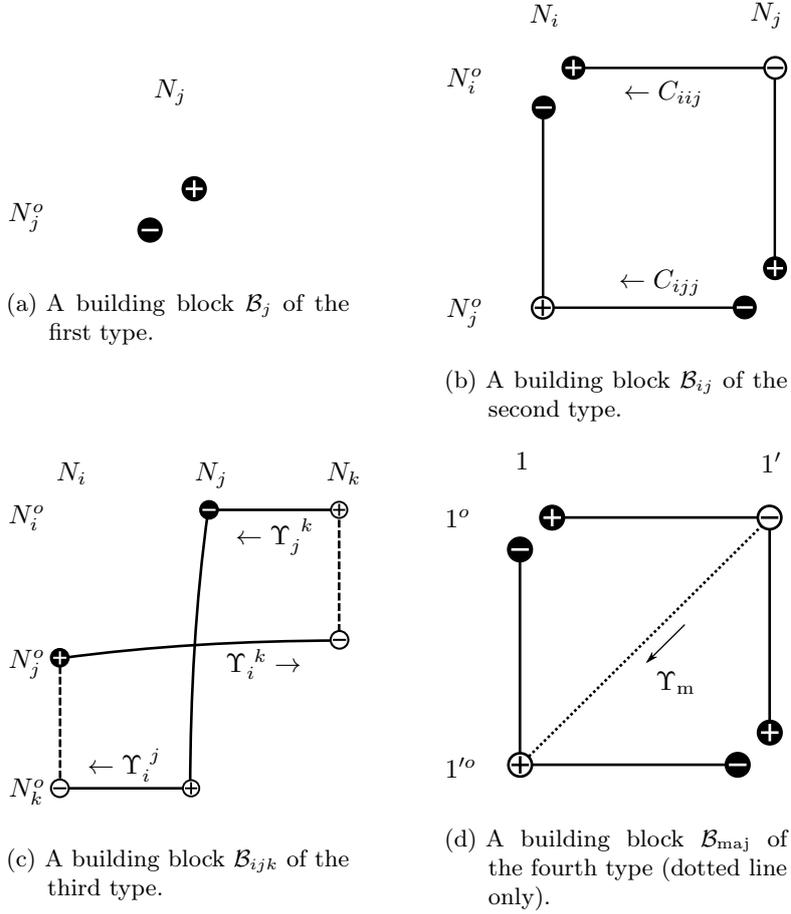

	\centering
	\begin{subfigure}{.3\textwidth}
		\centering
		\def\svgwidth{\textwidth}
		\subimport{\the\svgpath}{bb1.pdf_tex}
		\caption{A building block \B{j} of the first type.}
		\label{fig:bb1}
	\end{subfigure}
	\hspace{30pt}
\begin{subfigure}{.3\textwidth}
		\centering
		\def\svgwidth{\textwidth}
		\subimport{\the\svgpath}{bb2_Rplus.pdf_tex}
		\caption{A building block \B{ij} of the second type.}
		\label{fig:bb2}
	\end{subfigure}
	\\[10pt]
	\begin{subfigure}{.3\textwidth}
		\centering
		\def\svgwidth{\textwidth}
		\subimport{\the\svgpath}{bb3.pdf_tex}
		\caption{A building block \B{ijk} of the third type.}
		\label{fig:bb3}
	\end{subfigure}
	\hspace{30pt}
\begin{subfigure}{.3\textwidth}
		\centering
		\def\svgwidth{\textwidth}
		\subimport{\the\svgpath}{bb4.pdf_tex}
		\caption{A building block \BBBB{11'} of the fourth type (dotted line only).}
		\label{fig:bb4}
	\end{subfigure}

\captionsetup{width=.9\textwidth}
\caption{The Krajewski diagrams of four of the five different building blocks of almost-commutative geometries that potentially yield supersymmetric actions. The first corresponds to a gaugino--gauge boson pair, the second to a fermion--sfermion pair, the third to a superpotential interaction and the fourth to a Majorana mass for a gauge singlet. The fifth building block (not shown here) corresponds to a mass-like term for two fermions in the same representation of the gauge group. Note that the third block only contains the edges, not the vertices. The fourth block only contains the dotted edge. The sign inside the vertices represents their chirality. The white vertices correspond to fermions that have $R$-parity equal to $1$, i.e.~that are SM-like (and can consequently come in multiplicities higher than one). The black vertices have $R$-parity $-1$ and correspond to superpartners of bosons such as those of the SM.}
\label{fig:bbs}
\end{figure}

Thus, \emph{given an almost-commutative geometry}, the corresponding action is fixed. When we are talking about supersymmetric almost-commutative geometries, we mean those whose action \eqref{eq:totalaction} is \emph{supersymmetric}, i.e.
\begin{align}\label{eq:susy_action_no_aux}
	\delta S[\sfer, \psi, \gau{}, A] &:=  \frac{\mathrm{d}}{\mathrm{d} t} S[\sfer + t\delta\sfer, \psi + t\delta \psi, \gau{} + t\delta\gau{}, A + t\delta A]\Big|_{t = 0} = 0.
\end{align} 
Here with $\sfer$, $\psi$, $\gau{}$ and $A$ we generically denote the respective \emph{sfermions} and \emph{Higgs scalars}, \emph{fermions} and \emph{higgsinos}, \emph{gauginos} and \emph{gauge bosons} of the theory and $\delta \zeta$ ($\zeta = \sfer, \psi, \gau{}, A$) are the \emph{supersymmetry transformations} that feature the superpartner of the respective field.\footnote{Equation \ref{eq:susy_action_no_aux} does not mention the auxiliary fields that are needed to ensure supersymmetry both on shell and off shell. The spectral action gives an on shell action and needs to be written off shell, introducing auxiliary fields $D$ and $F$. These then both appear in the action and in the supersymmetry transformations \cite{BS13I}.} The central result of \cite{BS13I} was that each non-commutative geometry that is fully decomposable in the five \emph{building blocks} \B{i}, \Bc{ij}{\pm}, \B{ijk}, \B{\mathrm{maj}} and \B{\mathrm{mass}} (the first four of which are depicted in Figure \ref{fig:bbs}) are eligible to have a supersymmetric action and do so when they satisfy certain additional demands (see \cite[\S 3]{BS13I}). Here, the building block \B{i} describes a gaugino--gauge boson pair in the adjoint representation of $SU(N_i)$ and corresponds to a vector multiplet in the parlance of superfields. The building block \Bc{ij}{\pm} of the second type (which requires building blocks \B{i}, \B{j} of the first type) describes a fermion--sfermion pair in the representation \rep{i}{j} with the fermion $\fer{ij}$ being left-handed ($+$) or right-handed ($-$), respectively. This corresponds to a chiral multiplet. The third building block \B{ijk} (requiring building blocks \Bc{ij}{\pm}, \Bc{ik}{\mp} and \Bc{jk}{\pm} of the second type) describes extra fermionic and bosonic interactions and corresponds to a term in a superpotential consisting of the product of three different chiral multiplets. The building block $\B{\mathrm{maj}}$ (requiring a singlet \B{11'}) corresponds to a Majorana mass for a gauge singlet. Finally the fifth building block $\B{\mathrm{mass}}$ (not depicted in Figure \ref{fig:bbs}, requiring two building blocks \B{ij} of opposite chirality) describes a mass-like term between two different fermions in the same representation.\\

These building blocks do not automatically imply that the corresponding action is also supersymmetric: we have come across a number of possible obstacles for a supersymmetric action. These are the following:

\begin{itemize}


\item the three obstructions from Remark 9, Remark 18 and Proposition 24 of \cite{BS13I} concerning the set up of the almost-commutative geometry. The first excludes a finite algebra that is equal to $\com$ with the corresponding building block \B{1}, since it lacks gauge interactions and thus cannot be supersymmetric. The second excludes a finite algebra consisting of two summands that are both matrix algebras over $\com$ in the presence of only building blocks of the second type whose off-diagonal representations in the Hilbert space have $R$-parity equal to $-1$. The third obstruction says that for an algebra consisting of three or more summands $M_{N_{i,j,k}}(\com)$ we cannot have two building blocks \B{ij} and \B{ik} of the second type that share one of their indices. To avoid this obstruction, we can maximally have two components of the algebra that are a matrix algebra over $\com$.
%


\item to obtain the fermion--sfermion--gaugino interactions needed for a supersymmetric action, the parameters $C_{iij}$ and $C_{ijj}$ of the finite Dirac operator associated to a building block \B{ij} of the second type ---that read $\Cw{i,j}$ and $\Cw{j,i}$ after normalizing the kinetic terms of the sfermions--- should satisfy
\ba\label{eq:bb2-resultCiij}
	\Cw{i,j} &= \sgnc_{i,j} \sqrt{\frac{2}{\K_i}}g_i \id_M, &
	\Cw{j,i} &= \sgnc_{j,i} \sqrt{\frac{2}{\K_j}}g_j \id_M.
\ea	
Here $\sgnc_{i,j}$ and $\sgnc_{j,i}$ are signs that we are free to choose. The $\K_{i,j}$ are the pre-factors of the kinetic terms of the gauge bosons that correspond to the building blocks \B{i,j} of the first type and should be set to $1$ to give normalized kinetic terms (the consequences of this will be reviewed at the end of Section \ref{sec:identification}). The $g_{i,j}$ are coupling constants. Furthermore, these variables should act trivially on \emph{family space} (consisting of $M$ \emph{generations}), indicated by the identity $\id_M$ on family space. Similarly, when a building block \B{ijk} of the third type is present, its fermionic interactions can only be part of a supersymmetric action if the parameters $\yuk{i}{j}$, $\yuk{i}{k}$ and $\yuk{j}{k}$ of the finite Dirac operator satisfy 
\ba
\yuk{j}{k}C_{jkk}^{-1} &= - (C_{ikk}^*)^{-1}\yuk{i}{k}, &
(C_{iik}^*)^{-1}\yuk{i}{k} &= - \yuk{i}{j}C_{iij}^{-1}, &
\yuk{i}{j}C_{ijj}^{-1} &= - \yuk{j}{k}C_{jjk}^{-1}.\label{eq:improvedUpsilons}
\ea
For any building block of the third type it is necessary that either one or all three representations \rep{i}{j}, \rep{i}{k} and \rep{j}{k} in the Hilbert space have $R$-parity $-1$. The above relation assumes \rep{i}{j} to have $R = -1$, but the identities for the other cases are very similar \cite[\S 2.3]{BS13I}.


\item for the four-scalar interactions to have an off shell counterpart that satisfies the constraints supersymmetry puts on them, the coefficients of the interactions with the auxiliary fields $G_{i}$, $H$ and $F_{ij}$ should satisfy the demands listed in \cite[\S 3]{BS13I}.

\end{itemize}
For each almost-commutative geometry that one defines in terms of the building blocks, we should explicitly check that the obstructions are avoided and the appropriate demands are satisfied.



In the next section we will list the basic properties of the almost-commutative geometry that is to give the MSSM, including the building blocks it consists of and show that this set up avoids the three possible obstructions from the first item in the list above. To confirm that we are on the right track we identify all MSSM particles and examine their properties in Section \ref{sec:identification}. Finally, in Section \ref{sec:ncmssm-checks} we will confront our model with the demands from the last item in the list above. Throughout this paper, we will a priori allow for a number of generations other than $3$.

\section{The building blocks of the MSSM}\label{sec:NCSSM}
We start by listing the properties of the finite spectral triple that, when part of an almost-commutative geometry, should correspond to the MSSM.

\begin{figure}[ht!]
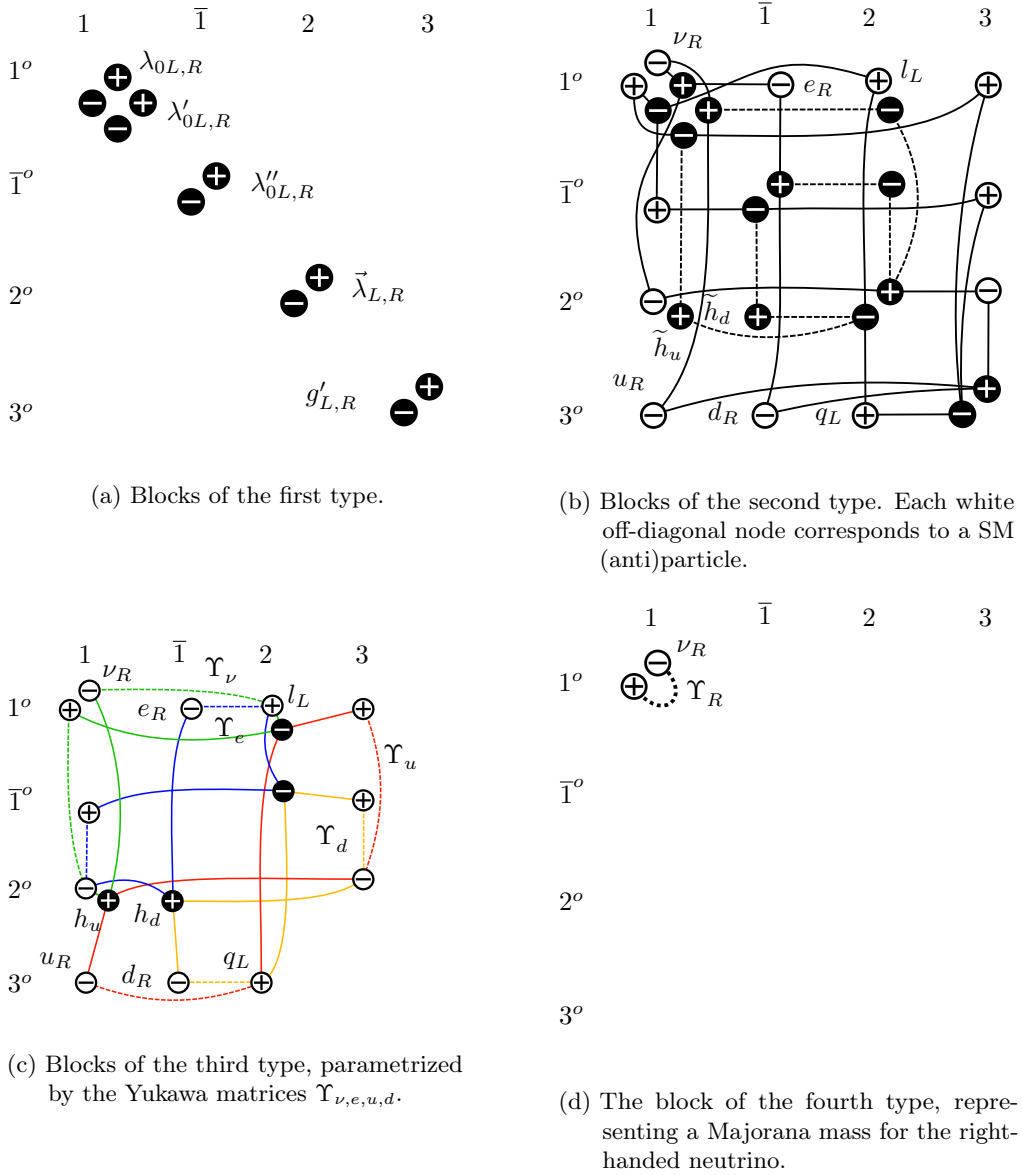

	\centering
	\begin{subfigure}{.4\textwidth}
		\centering
		\def\svgwidth{\textwidth}
		\subimport{\the\svgpath}{MSSM_bb1_compact.pdf_tex}
		\caption{Blocks of the first type.\\\ \\\ }
		\label{fig:MSSM_bb1_compact}
	\end{subfigure}
	\hspace{30pt}
\begin{subfigure}{.4\textwidth}
		\centering
		\def\svgwidth{\textwidth}
		\subimport{\the\svgpath}{MSSM_bb2_compact.pdf_tex}
		\caption{Blocks of the second type. Each white off-diagonal node corresponds to a SM (anti)particle.}
		\label{fig:MSSM_bb2_compact}
	\end{subfigure}
	\\[10pt]
	\begin{subfigure}{.4\textwidth}
		\centering
		\def\svgwidth{\textwidth}
		\subimport{\the\svgpath}{MSSM_bb3_compact2.pdf_tex}
		\caption{Blocks of the third type, parametrized by the Yukawa matrices $\Upsilon_{\nu, e, u, d}$.\\\ }
		\label{fig:MSSM_bb3_compact}
	\end{subfigure}
	\hspace{30pt}
\begin{subfigure}{.4\textwidth}
		\centering
		\def\svgwidth{\textwidth}
		\subimport{\the\svgpath}{MSSM_bb4_compact2.pdf_tex}
		\caption{The block of the fourth type, representing a Majorana mass for the right-handed neutrino.}
		\label{fig:MSSM_bb4_compact}
	\end{subfigure}
\captionsetup{width=.9\textwidth}
\caption{All building blocks that together represent the particle content and interactions of the MSSM.}
\label{fig:MSSM-bbs}
\end{figure}

\begin{enumerate}


\item The gauge group of the MSSM is (up to a finite group) the same as that of the SM. In noncommutative geometry there is a strong connection between the algebra $\A$ of the almost-commutative geometry and the gauge group $\mathcal{G}$ of the corresponding theory. There is more than one algebra that may yield the correct gauge group (Lemma 1 of \cite{BS12}) but any supersymmetric extension of the SM also contains the SM particles, which requires an algebra that has the right representations (see just below the aforementioned Lemma). This motivates us to take the Standard Model algebra:
	\begin{align}
		\A_F \equiv \A_{SM} = \com \oplus \mathbb{H} \oplus M_3(\com).\label{eq:ncmssm-alg}
	\end{align}
Note that with this choice we already avoid the third obstruction for a supersymmetric theory from the first item in the list above, since only two of the summands of this algebra are defined over $\com$.

In the derivation \cite{CCM07} of the SM from noncommutative geometry the authors first start with the `proto-algebra'
\ba
	\A_{L,R} = \com \oplus \mathbb{H}_L \oplus \mathbb{H}_R \oplus M_3(\com)\label{eq:LRalgebra}
\ea
(cf.~\cite[\S2.1]{CCM07}) that breaks into the algebra above after allowing for a Majorana mass for the right-handed neutrino \cite[\S 2.4]{CCM07}. Although we do not follow this approach here, we do mention that this algebra avoids the same obstruction too.

\item As is the case in the NCSM, we allow four inequivalent representations of the components of \eqref{eq:ncmssm-alg}: $\mathbf{1}$, $\overline{\mathbf{1}}$, $\mathbf{2}$ and $\mathbf{3}$. Here $\overline{\mathbf{1}}$ denotes the real-linear representation $\pi(\lambda)v = \bar\lambda v$, for $ v\in\overline{\mathbf{1}}$.\footnote{Keep in mind that we ensure the Hilbert space being complex by defining it as a bimodule of the complexification $\A^{\com}$ of $\A$, rather than of $\A$ itself \cite{CC08}.} This results in only three independent forces ---with coupling constants $g_1$, $g_2$ and $g_3$--- since the inner fluctuations of the canonical Dirac operator acting on the representations $\mathbf{1}$ and $\overline{\mathbf{1}}$ of $\com$ are seen to generate only a single $u(1)$ gauge field \cite[\S 3.5.2]{CCM07} (see also Section \ref{sec:ncmssm_unimod}).

\item If we want a theory that contains the superpartners of the gauge bosons, we need to define the appropriate building blocks of the first type (cf.~\cite[\S 2.1]{BS13I}). In addition, we need these building blocks to define the superpartners of the various Standard Model particles. We introduce
 \ba \label{eq:mssm-bb1s}
\B{1},\quad \B{1_R},\quad \B{\bar 1_R},\quad \B{2_L},\quad \B{3},
\ea
whose representations in $\H_F$ all have $R = -1$ to ensure that the gauginos and gauge bosons are of opposite R-parity. The Krajewski diagram that corresponds to these building blocks is given in Figure \ref{fig:MSSM_bb1_compact}. For reasons that will become clear later on, we have two building blocks featuring the representation $\mathbf{1}$, and one featuring $\overline{\mathbf{1}}$. We distinguish the first two by giving one a subscript $R$. This notation is not related to $R$-parity but instead is inspired by the derivation of the Standard Model where, in terms of the proto-algebra \eqref{eq:LRalgebra}, the component $\com$ is embedded in the component $\mathbb{H}_R$ via $\lambda \to \diag(\lambda, \bar\lambda)$. The initially two-dimensional representation $\mathbf{2}_R$ of this component (making the right-handed leptons and quarks doublets) thus breaks up into two one-dimensional representations $\mathbf{1}_R$ and $\mathbf{\bar 1}_R$ (corresponding to right-handed singlets).

At this point we thus have too many fermionic degrees of freedom, but these will be naturally identified to each other in Section \ref{sec:identification}.

\item For each of the Standard Model fermions\footnote{In the strict sense the Standard Model does not feature a right handed neutrino (nor does the MSSM), but allows for extensions that do. On the other hand the more recent derivations of the SM from noncommutative geometry naturally come with a right-handed neutrino. We will incorporate it from the outset, always having the possibility to discard it should we need to.} we define the corresponding building block of the second type:
	\begin{subequations}\label{eq:mssm-bb2s}
	\begin{align}
		\Bc{1_R1}{-} &: (\nu_R, \widetilde{\nu}_R), & 	\Bc{\bar 1_R1}{-} &: (e_R, \widetilde{e}_R), & \Bc{2_L1}{+} &: (l_L, \widetilde{l}_L), \\
		\Bc{1_R3}{-} &: (u_R, \widetilde{u}_R), & 	\Bc{\bar 1_R3}{-} &: (d_R, \widetilde{d}_R), & \Bc{2_L3}{+} &: (q_L, \widetilde{q}_L).
\intertext{Of each of the representations in the finite Hilbert space we will take $M$ copies representing the $M$ generations of particles, also leading to $M$ copies of the sfermions. We can always take $M = 3$ in particular. Each of these fermions has $R = +1$. We do the same for representations in which the SM Higgs resides:}
		\B{1_R2_L} &: (h_u, \widetilde{h}_u), & 	\B{\bar 1_R2_L} &: (h_d, \widetilde{h}_d), &  &
	\end{align} 
\end{subequations}
save that their representations in the Hilbert space have $R = -1$ and consequently we take only one copy of both. For the two Higgs/higgsino building blocks we can choose the grading still. We will set them both to be left-handed and justify that choice later. 

The Krajewski diagram that corresponds to these building blocks is given by Figure \ref{fig:MSSM_bb2_compact}. 

	The fact that there is at least one building block \B{1j}, $j = \bar 1_R, 2_L, 3$, avoids the first of the three obstructions for a supersymmetric theory mentioned in the first item of the list above. 

The building blocks introduced above fully determine the finite Hilbert space. For concreteness, it is given by
\ba
	\H_F &= \H_{F, R = +} \oplus \H_{F, R = -}, \label{eq:ncmssm_H_F}
\ea
with $\H_{F, R = \pm} = \tfrac{1}{2}(1 \pm R)\H_F$ (cf.~\cite[\S 1.5]{BS13I}) reading
\bas
			\H_{F, R = +} &= \big(\mathcal{E} \oplus \mathcal{E}^o\big)^{\oplus M},& \mathcal{E} &= (\mathbf{2}_L \oplus \mathbf{1}_R \oplus \overline{\mathbf{1}}_R) \otimes (\mathbf{1} \oplus \mathbf{3})^o\nn\\
			\H_{F, R = -} &= \mathcal{F} \oplus \mathcal{F}^o, &  \mathcal{F} &= (\mathbf{1} \otimes \mathbf{1}^o)^{\oplus 2}\ \oplus\ \overline{\mathbf{1}}\otimes \overline{\mathbf{1}}^o\ \oplus\ \mathbf{2} \otimes \mathbf{2}^o\ \nn\\
		&& &\qquad\oplus\ \mathbf{3} \otimes \mathbf{3}^o \oplus\ (\mathbf{1}_R \oplus \overline{\mathbf{1}}_R) \otimes \mathbf{2}^o_L.\nn
\eas
Here $\mathcal{E}$ contains the finite part of the left- and right-handed leptons and quarks. The first four terms of $\mathcal{F}$ represent the $u(1)$, $su(2)$ and $su(3)$ gauginos and the last term the higgsinos. For the (MS)SM the number of generations $M$ is equal to $3$.

\item In terms of the `proto-algebra' \eqref{eq:LRalgebra} the operator 
\bas
	R = - (+, -, -, +) \otimes (+, -, -, +)^o
\eas
gives the right values for $R$-parity to all the fermions: $R = + 1$ for all the SM-fermions, $R = -1$ for the higgsino-representations that are in $\mathbf{2}_R\otimes \mathbf{2}_L^o$ before breaking to $(\mathbf{1}_R \oplus \overline{\mathbf{1}}_R) \otimes \mathbf{2}_L^o$.

Since there is at least one building block of the second type whose representation in the finite Hilbert space has $R = +1$, also the second obstruction for a supersymmetric theory mentioned above is avoided. 

\item The MSSM features additional interactions, such as the Yukawa couplings of fermions with the Higgs. In the superfield formalism, these are determined by a superpotential. Its counterpart in the language of noncommutative geometry is given by the building blocks \BBB{ijk} of the third type. These should at least contain the Higgs-interactions of the Standard Model (but with the distinction between up- and down-type Higgses). The values of the grading on the representations in the finite Hilbert space are such that they allow us to extend the Higgs-interactions to the following building blocks:
\begin{align}\label{eq:mssm-bb3s}
\BBB{11_R2_L},&&	\BBB{1\bar 1_R2_L},&& \BBB{1_R2_L3},&& \BBB{\bar 1_R2_L3}. 
\end{align}
The four building blocks \BBB{ijk}, are depicted in Figure \ref{fig:MSSM_bb3_compact}. (For conciseness we have omitted here the building blocks of the first type and the components of $D_F$ from the building blocks of the second type.) 

Note that all components of $D_-$, the part of $D_F$ that anticommutes with $R$, that are allowed by the principles of NCG are in fact also non-zero now. This is in contrast with those of $D_+$, on which the (ad hoc) requirement \cite[\S 2.6]{CCM07} to commute with
\bas
	\com_F :=\{ (\lambda, \diag(\lambda, \bar\lambda), 0), \lambda \in \com\} \subset \A_{SM}
\eas
is imposed. The reason for this is to keep the photon massless and to get the interactions of the SM. Requiring the same for the entire finite Dirac operator would forbid the majority of the components that determine the sfermions, not requiring it at all would lead to extra, non-supersymmetric interactions such as $\repl{\bar1}{1} \to \repl{3}{1}$. 
Thus, we slightly change the demand, reading
\ba
	[D_+, \com_F] = 0.\label{eq:demandDF}
\ea
Relaxing this demand does not lead to a photon mass since it only affects the sfermions that have $R = -1$ whereas any photon mass would arise from the kinetic term of the Higgses, having $R = +1$.

At this point we can justify the choice for the grading of the up- and down-type higgsinos. If the grading of any of the two would have been of opposite sign, none of the building blocks of the third type that feature that particular higgsino could have been defined. The interactions that are still possible then cannot be combined into building blocks of the third type, which is an undesirable property. It corresponds to a superpotential that is not holomorphic (see \cite[\S 2.3]{BS13I}). 


\item Having a right-handed neutrino in $\repl{1_R}{1}$, that is a singlet of the gauge group, we are allowed to add a Majorana mass for it via 
\ba\label{eq:mssm-bb4}\BBBB{11_R}\ea 
such as in \cite[\S 2.5.1]{BS13I}. This is represented by the dotted diagonal line in Figure \ref{fig:MSSM_bb4_compact}. The building block is parametrized by a symmetric $M\!\times\!M$--matrix $\yuk{R}{}$. 


\end{enumerate}

Summarizing things, the finite spectral triple of the almost-commutative geometry that should yield the MSSM then reads
\ba
	& \B{1} \oplus \B{1_R} \oplus \B{\bar1_R} \oplus \B{2} \oplus \B{3} \oplus \Bc{1_R2_L}{+} \oplus \Bc{\bar1_R2_L}{+}\nn\\
	&\qquad\oplus \Bc{1_R1}{-} \oplus \Bc{\bar1_R1}{-} \oplus \Bc{2_L1}{+} \oplus \Bc{1_R3}{-} \oplus \Bc{\bar1_R3}{-} \oplus \Bc{2_L3}{+}\nn\\
	&\qquad\qquad\oplus \B{11_R2_L} \oplus \B{1\bar1_R2_L} \oplus \B{1_R2_L3} \oplus \B{\bar1_R2_L3} \oplus \B{\textrm{maj}}\label{eq:ncmssm-acg}
\ea
One of its properties is that all components that are not forbidden by the principles of NCG and the additional demand \eqref{eq:demandDF} are in fact also non-zero, save for the supersymmetry-breaking gaugino masses \cite{BS13II} that we will not cover here. 

\begin{figure}[ht]
	\centering
	\centering
	\def\svgwidth{.4\textwidth}
		\subimport{\the\svgpath}{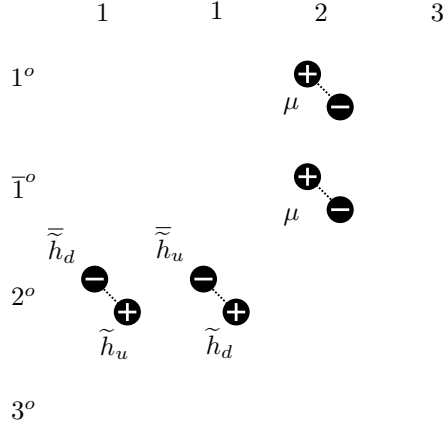}
		\captionsetup{width=.6\textwidth}
		\caption{The extra building blocks of the second type featuring a Higgs/higgsino-pair and the building blocks of the fifth type that are consequently possible.}
		\label{fig:MSSM_bb5_compact}
\end{figure}

\begin{rmk}\label{rmk:extra-higgsinos}
	Running ahead of things a bit already we note that there is an important difference with the MSSM. In the superfield-formalism there is an interaction that reads
\ba\label{eq:mu-term}
	\mu H_d \cdot H_u,
\ea
where $H_{u,d}$ represent the up-/down-type Higgs/higgsino superfields \cite[\S 8.3]{DGR04}. Suppose that \Bc{1_R2_L}{+} and \Bc{\bar1_R2_L}{+} indeed describe the up- and down-type Higgses and higgsinos. Because their vertices are on different places in the Krajewski diagram and in addition they have the same value for the grading, there is no building block of the fifth type possible that would be the equivalent of \eqref{eq:mu-term}. Moreover, in the MSSM there is a soft supersymmetry-breaking interaction
\bas
	B\mu h_d\cdot h_u + h.c.
\eas
In this framework also such an interaction can only be generated via a building block of the fifth type (in combination with gaugino masses, see \cite[\S 4.4]{BS13II}). Not having these interactions would at least leave several of the tree-level mass-eigenstates that involve the Higgses massless \cite[\S 10.3]{DGR04}. We can overcome this problem by adding two more building blocks \B{1_R2_L} and \B{\bar1_R2_L} of the second type whose values of the grading are opposite to the ones previously defined. With these values no additional components for the finite Dirac operator are possible, except for two building blocks of the fifth type that run between the representations of \Bc{1_R2_L}{\pm} and between those of \Bc{\bar1_R2_L}{\pm}. If we then identify the degrees of freedom of \Bc{1_R2_L}{+} to those of \Bc{\bar1_R2_L}{-} and those of \Bc{\bar1_R2_L}{+} to those of \Bc{1_R2_L}{-}, this would give us the interactions that correspond to the term \eqref{eq:mu-term}. The additions to the finite spectral triple \eqref{eq:ncmssm-acg} that correspond to these steps are given by
\ba\label{eq:ncmssm-acg-ext}
	\Bc{1_R2_L}{-} \oplus \Bc{\bar 1_R2_L}{-} \oplus \B{\mathrm{mass}, 1_R2_L} \oplus \B{\mathrm{mass}, \bar1_R2_L}.
\ea
This situation is depicted in Figure \ref{fig:MSSM_bb5_compact}.
\end{rmk}

We proceed by ensuring that we are indeed talking about the noncommutative counterpart of the MSSM by identifying the MSSM particles and checking that the number of fermionic and bosonic degrees of freedom are the same.

\section{Identification of particles and sparticles}\label{sec:identification}

\subsection{The gauge group and hypercharges}

To justify the nomenclature we have been using in the previous section we need to test the properties of the new particles by examining how they transform under the gauge group (e.g.~\cite[\S 7.1]{V06}). We do this by transforming elements of the finite Hilbert space and finite Dirac operator under the gauge group according to 
\begin{align*}
	\H_F \ni \psi&\to U\psi,	& D_F &\to UD_FU^*, 
\end{align*}
with $U = uJuJ^*$, $u \in SU(\A)$, but with a definition of the gauge group featuring the $R$-parity operator:
\begin{align*}
	SU(\A) := \{ u \in \A\ |\ uu^* = u^*u = 1, \det{}_{\H_{F, R = +}}(u) = 1\}.
\end{align*}	
(See the discussion in Section 1.5 of \cite{BS13I}.)
Since we have $\H_{F, R = +} = \H_{F, SM}$, the space that describes the SM fermions, this determinant gives
\ba\label{eq:sm_gauge_group}
	SU(\A_{SM}) = \{(\lambda, q, m) \in U(1) \times SU(2) \times U(3), [\lambda\det(m)]^{4M} = 1\}.
\ea
The factor $M$ again represents the number of particle generations and stems from the fact that the algebra acts trivially on family-space. Unitary quaternions $q$ automatically have determinant $1$ and consequently all contributions to the determinant come from 
\bas
	\mathcal{E}^o = (\mathbf{1} \oplus \mathbf{3})\otimes (\mathbf{2}_L \oplus \mathbf{1}_R \oplus \overline{\mathbf{1}}_R)^o
\eas 
defined above, instead of from $\mathcal{E}$. The power $4 = 2 + 1 + 1$ above comes from the second part of the tensor product on which the unitary elements $U(\A)$ act trivially. From \eqref{eq:sm_gauge_group} we infer that the $U(1)$-part of $SU(\A_{SM})$ (the part that commutes with all other elements) is given by 
\begin{align}
	\{(\lambda, 1, \lambda^{-1/3}1_3), \lambda \in U(1)\} \subset SU(\A_{SM}).\label{eq:U1factor}
\end{align}
This part determines the hypercharges of the particles; these are given by the power with which $\lambda$ acts on the corresponding representations. This result makes the identification of the fermions that have $R = +1$ exactly the same as in the case of the SM (\cite[\S 2.5]{CCM07}).
Applying it to the gaugino and higgsino sectors of the Hilbert space, we find that:
\begin{itemize}
	\item there are the gauginos $\widetilde{g} \in \repl{3}{3}$ whose traceless part transforms as $\widetilde{g} \to  \bar v\widetilde{g} v^t$, with $\bar v\in SU(3)$ (i.e.~it is in the adjoint representation of $SU(3)$) and whose trace part transforms trivially;
	\item there are the gauginos $\widetilde{W} \in \repl{2}{2}$ whose traceless part transforms according to $\widetilde{W} \to q\widetilde{W}q^*$ with $q \in SU(2)$ (i.e.~the adjoint representation of $SU(2)$) and whose trace part transforms trivially;
	\item the higgsinos in \repl{1_R}{2_L} and \repl{\bar1_R}{2_L} transform in the representation $\mathbf{2}$ of $SU(2)$ and have hypercharge $+1$ and $-1$ respectively;
	\item the gauginos in \repl{1}{1}, \repl{2}{2} and \repl{3}{3} all have zero hypercharge.
\end{itemize}
The new scalars, parametrized by the finite Dirac operator, generically transform as $\Phi \to U\Phi U^*$. In particular, we separately consider the elements $U = uJuJ^*$ with $u = (\lambda, 1, \lambda^{-1/3}1_3)$, $(1, q, 1)$ and $(1, 1, \bar v)$. This gives the following:  
\begin{itemize}
\item with $u = (\lambda, 1, \lambda^{-1/3}1_3)$ we find for the hypercharges of the various sfermions: 
\begin{align*}
	\widetilde{q}_L&:\quad \tfrac{1}{3}, & \widetilde{u}_R &:\quad \tfrac{4}{3},& \widetilde{d}_R &:\quad -\tfrac{2}{3},\\
	\widetilde{l}_L&:\quad -1, & \widetilde{\nu}_R &:\quad 0,& \widetilde{e}_R &:\quad -2.
\end{align*}
The conjugates are found to carry the opposite charge.
\item with $u = (1, q, 1)$ we find the following sfermions that transform non-trivially: $\widetilde{q}_L$ and $\widetilde{l}_L$, each coming in $M$ generations.
\item with $u = (1, 1, \overline{v})$ we find the following sfermions that transform in the fundamental representation of $SU(3)$: $\widetilde{q}_L$, $\widetilde{u}_R$ and $\widetilde{d}_R$, each coming in $M$ generations. 
\end{itemize}

This completes the identification of the new elements in the theory with the gauginos, higgsinos and sfermions of the MSSM. 

\subsection{Unimodularity in the MSSM}\label{sec:ncmssm_unimod}

Having identified the particles there is one other thing to check; that the number of bosonic and fermionic degrees of freedom are indeed the same. We can quite easily see that at least initially this is not the case for the following reason. In order to be able to define the building blocks \Bc{\overline{1}_R1}{-}, \Bc{\overline{1}_R3}{-} and \Bc{\overline{1}_R2_L}{+} of the second type (describing the right-handed (s)electron and (s)quark and down-type Higgs/higgsino respectively), we defined the building blocks \B{\overline{1}} and \B{1_R} of the first type. Each provides extra $u(1)$ fermionic degrees of freedom, but no bosonic ones (see below). In addition, the gaugino $\widetilde{W}$ contains a trace part, whereas the corresponding gauge boson does not.

We will employ the unimodularity condition 
\ba \label{eq:unimod-R} \tr_{\H_{F, R = +}}A_\mu = 0\ea to reduce the bosonic degrees of freedom on the one hand and see what its consequences are, using the supersymmetry transformations.\\

First of all, we note that the inner fluctuations on the $\mathbf{1}$ and $\overline{\mathbf{1}}$ give rise to only one $u(1)$ gauge field (cf.~\cite[\S 15.4]{CCM07}). Initially there are
\bas
	\Lambda &= i\gamma^\mu\sum_j  \lambda_j\partial_\mu \lambda_j',&& \text{and} &
	\Lambda' &= i\gamma^\mu \sum_j \bar\lambda_j\partial_\mu \bar\lambda_j',
\eas
but since $\Lambda$ must be self-adjoint (as $\dirac$ is), $\Lambda_\mu = i\sum_j  \lambda_j\partial_\mu \lambda_j'$ is real-valued. Consequently $\Lambda'_\mu(x) = - \Lambda_\mu(x)$ and they indeed generate the same gauge field. But via the supersymmetry transformations this also means that 
\bas
	\delta \Lambda \propto \delta \Lambda',
\eas 
i.e.~the corresponding gauginos whose finite parts are in $\mathbf{1} \otimes \mathbf{1}^o$ and $\overline{\mathbf{1}} \otimes \overline{\mathbf{1}}^o$ should be associated to each other.

Second, the inner fluctuations of the quaternions $\mathbb{H}$ generate an $su(2)$-valued gauge field. This can be seen as follows. The quaternions form a real algebra, spanned by $\{1_2, i\sigma^a\}$, with $\sigma^a$ the Pauli matrices. Since $\dirac$ commutes with the basis elements, the inner fluctuations
\bas
	\sum_j q_j [\dirac, q_j'],\qquad\ q_j, q_j' \in C^{\infty}(M, \mathbb{H})
\eas 
can again be written as a quaternion-valued function, i.e.~of the form
\bas
	 \sum_j f_{j0}[\dirac f_{j0}'] + f_{ja} [\dirac, if'_{ja}\sigma^a]
\eas
for certain $f_{j0}, f_{j0}', f_{ja}, f_{ja}' \in C^\infty(M, \mathbb{R})$.
Using that $[\dirac, x]^* = - [\dirac, x^*]$, only the second term above, which we will denote with $Q$, 
is seen to satisfy the demand of self-adjointness for the Dirac operator. Since the Pauli matrices are traceless, the self-adjoint inner fluctuations of $\mathbb{H}$ are automatically traceless as well.

Using the supersymmetry transformations on the gauge field $Q$, we demand that $\tr \delta Q = 0$, which sets the trace of the corresponding gaugino and auxiliary field equal to zero.  

Third, the inner fluctuations of the component $M_3(\com)$ of the algebra generate a gauge field
\bas
	V' = \sum_j m_j [\slashed{\partial}, m_j'],\qquad m_j, m_j' \in M_3(\com).
\eas
Because $D_A$ is self-adjoint $V'$ must be too and hence $V'(x) \in u(3)$. We can employ the unimodularity condition \eqref{eq:unimod-R}, which for $\H_F$ given by \eqref{eq:ncmssm_H_F} reads 
\bas
	4M(\Lambda + \tr V') = 0.
\eas
The contributions to this expression again only come from $\mathcal{E}^o$ and the factor $4 = 2 + 1 + 1$ arises from the gauge fields acting trivially on the second part of its tensor product. The inner fluctuations of the quaternions do not appear in this expression, since they are traceless. A solution to the demand above is
	\ba 
		V' = - V - \frac{1}{3}\Lambda \id_3,\label{eq:ident_V}
	\ea 
with $V(x) \in su(3)$. The sign of $V$ is chosen such that the interactions match those of the Standard Model \cite[\S 3.5]{CCM07}.

In order to introduce coupling constants into the theory, we have to redefine the fields at hand:
\bas
	\Lambda_\mu &\equiv g_1B_\mu, &
	Q_\mu &\equiv g_2W_\mu, &
	V_\mu &\equiv g_3g_\mu.
\eas
Note that we parametrize the gauge fields differently than in \cite{CCM07}.  
Then looking at the supersymmetry transformation of $V'$, we infer that its superpartner, the $u(3)$ `gluino' $\gluino{L,R}'$ and corresponding auxiliary field $G_3'$ can also be separated into a trace part and a traceless part. We parametrize them similarly as
\ba
	\gluino{L,R}' &= \gluino{L,R} - \frac{1}{3}\bino{L,R} \id_3, & G_3' &= G_3 - \frac{1}{3} G_1 \id_3,\label{eq:su3-ident}
\ea
with $\bino{L,R}$ the superpartner of $B_\mu$ and $G_1$ the associated auxiliary field. 

The unimodularity condition reduced a bosonic degree of freedom. Employing it in combination with the supersymmetry transformations allowed us to reduce fermionic and auxiliary degrees of freedom as well. A similar result comes from $\mathbf{1}$ and $\overline{\mathbf{1}}$ generating the same gauge field. All in all we are left with three gauge fields, gauginos and corresponding auxiliary fields:
\bas
	\photon_\mu &\in C^{\infty}(M, u(1)),&  
\bino{L,R} &\in L^2(M, S\otimes u(1)), &
	G_1 &\in C^{\infty}(M, u(1)),\\
	\weak_\mu &\in C^{\infty}(M, su(2)),&  
	\wino{L,R} &\in L^2(M, S\otimes su(2)),& 	
	G_2 &\in C^{\infty}(M, su(2)),\\
	\gluon_\mu &\in C^{\infty}(M, su(3)),& 
	\gluino{L,R} &\in L^2(M, S\otimes su(3)), &
	G_3 &\in C^{\infty}(M, su(3)),
\eas
exactly as in the MSSM.

With the finite Hilbert space being determined by the building blocks of the first and second type, we can also obtain the relation between the coupling constants $g_1$, $g_2$ and $g_3$ that results from normalizing the kinetic terms of the gauge bosons, appearing in \eqref{eq:spectral_action_acg_flat}. The latter are of the form
\ba
	\frac{1}{4}\K_{j} \int_M F^{j\, a}_{\mu\nu} F^{j\,a\,\mu\nu},\qquad \K_j = \frac{f(0)}{3\pi^2}g_j^2n_j \Big(2N_j + \sum_k M_{jk} N_k\Big) \equiv \frac{r_j}{3}\Big(2N_j + \sum_k M_{jk} N_k\Big),\label{eq:ncmssm-ident-gauge-kin}
\ea
where the label $j$ denotes the type (i.e.~$u(1)$, $su(2)$ or $su(3)$) of gauge field and the index $a$ runs over the generators of the corresponding gauge group. The expressions for $\K_j$ include a factor $2$ that comes from summing over both particles and anti-particles. Its first term stems from a building block \B{j} of the first type and the other terms come from the building blocks \B{jk} of the second type, having multiplicity $M_{jk}$. The symbol $n_j$ comes from the normalization 
\bas
	\tr T^a_{j} T^b_{j} = n_{j}\delta^{ab}
\eas
of the gauge group generators $T^a_j$. For $su(2)$ and $su(3)$ these have the value $n_{2,3} = \tfrac{1}{2}$, for $u(1)$ we have $n_1 = 1$. In addition, each contribution to the kinetic term of the $u(1)$ gauge boson must be multiplied with the square of the hypercharge of the building block the contribution comes from. The contributions (see \cite[\S 4.3]{BS12}) from each representation to each kinetic term appearing in the MSSM are given in Table \ref{tab:contr_G}.
\begin{table}[ht]
\begin{tabularx}{\textwidth}{XlllllX}
\toprule
& Particle			& Representation	& $\K_1$ 									&$\K_2$	& $\K_3$	& \\[0pt]
\midrule
& \bino{L,R} 		& \repl{1}{1}			& 0 											& 0 		& 0 & \\[0pt]
& \wino{L,R} 		& \repl{2}{2}			& 0 											& 4			& 0 & \\[0pt]
& \gluino{L,R}	& \repl{3}{3} 		& 0 											& 0			& 6& \\[0pt]
\midrule
& $\nu_R$ 			& \repl{1}{1}			& $0$ 										& 0			& 0 & \\[0pt]
& $e_R$					& \repl{1}{\bar 1}& $4M$ 										& 0			& 0 & \\[0pt]
& $l_L$				& \repl{1}{2}			& $2M$ 										& $M$	& 0 & \\[0pt]
& $d_R$ 				& \repl{\bar 1}{3}&	$3(-1+\frac{1}{3})^2M$	& 0			& $M$ & \\[0pt]
& $u_R$					& \repl{1}{3}			&	$3(1+\frac{1}{3})^2M$		& 0			& $M$ & \\[0pt]
& $q_L$					& \repl{2}{3}			&	$6(\frac{1}{3})^2M$		& $3M$	& $2M$ & \\[0pt]
\midrule
& $\hd$ 				& \repl{\bar 1}{2}& 2 											& 1			& 0 & \\[0pt] 
& $\hu$ 				& \repl{1}{2}			& 2 											& 1			& 0 & \\[0pt]
\midrule
& Total 				& 								& $4+ 120M/9$			& $6 + 4M$ & $6 + 4M$ &\\[0pt]
\bottomrule
\end{tabularx}
\caption[The contributions to the pre-factors of the gauge bosons]{The contributions to the pre-factors \eqref{eq:ncmssm-ident-gauge-kin} of the gauge bosons' kinetic terms for all of the representations of the MSSM. The number of generations is denoted by $M$.}
\label{tab:contr_G}
\end{table}

Summing all contributions, we find  
\bas
	\K_1 &= \frac{f(0)}{3\pi^2}n_1g_1^2(4 + 120M/9) \equiv \frac{r_1}{3}(4 + 120M/9),\nn\\ 
	\K_2 &= \frac{f(0)}{3\pi^2}n_2 g_2^2(6 + 4M) \equiv \frac{r_2}{3}(6 + 4M),  \nn\\
	\K_3 &= \frac{f(0)}{3\pi^2}n_3 g_3^2(6 + 4M) \equiv \frac{r_3}{3}(6 + 4M),
\eas
for the coefficients of the gauge bosons' kinetic terms. We have to insert an extra factor $\tfrac{1}{4}$ into $\K_1$, since we must divide the hypercharges by two to compare with \cite{CCM07}, that has a different parametrization of the gauge fields.
Normalizing these kinetic term by setting $\K_{1,2,3} = 1$, we obtain for the $r_i$ (defined in \eqref{eq:ncmssm-ident-gauge-kin}):
\ba\label{eq:ncmssm-rs}
	r_3 &= r_2 = \frac{3}{6 + 4M},& r_1 &= \frac{9}{3 + 10M}.
\ea
Consequently, we find for the coefficients 
\ba\label{eq:defw}
		\w{ij} := 1 - r_iN_i - r_jN_j
\ea
the following values:
\bas
	\w{11} &= \frac{10M - 15}{10M + 3\phantom{1}},&
	\w{12} &= \frac{20 M^2 - 12M -27}{20 M^2 + 36M + 9\phantom{1}},\\
	\w{13} &= \frac{40 M^2 - 54 M - 63}{40 M^2 + 72 M + 18}, &
	\w{23} &= \frac{4M - 9}{4M + 6}.
\eas

From \eqref{eq:ncmssm-rs} it is immediate that, upon taking $M = 3$ and inserting the values of $n_{1,2,3}$, the three coupling constants are related by
\ba
	g_3^2 = g_2^2 = \frac{11}{9} g_1^2.\label{eq:mssm-gut}
\ea
This is different than for the SM \cite[\S 4.2]{CCM07}, where it is the well-known $g_2^2 = g_3^2 = \tfrac{5}{3}g_1^2$. For this value of $M$, the $\w{ij}$ have the following values:
\ba
	\w{11} &= \frac{5}{11},&
	\w{12} &= \frac{13}{33}, &
	\w{13} &= \frac{5}{22}, &
	\w{23} &= \frac{1}{6}.\label{eq:values-omega-3gen}
\ea

\begin{rmk}
	In Remark \ref{rmk:extra-higgsinos} we have suggested to add one extra copy of the two building blocks that describe the Higgses and higgsinos, to match the interactions of the MSSM. Such an extension gives extra contributions to the kinetic terms of the $su(2)$ and $u(1)$ gauge bosons, leading to
	\ba\label{eq:ncmssm-rs2}
		r_3 &= \frac{3}{6 + 4M},&
		r_2 &= \frac{3}{8 + 4M},&
		r_1 &= \frac{9}{6 + 10M}.&
	\ea
Consequently,
\bas
	\w{11} &= \frac{5M - 6}{5M + 3},&
	\w{12} &= \frac{10 M^2 + 2M -15}{2 (2 + M) (3 + 5 M)},\\
	\w{13} &= \frac{20 M^2 - 21 M - 36}{2 (3 + 2 M) (3 + 5 M)}, & 
	\w{23} &= \frac{4 M^2 - M - 15}{2(2 + M) (3 + 2 M)}
\eas
for the parameters $\w{ij}$. From the ratios of the $r_1$, $r_2$ and $r_3$ we derive for the coupling constants when $M = 3$:
\bas
	g_3^2 &= \frac{10}{9}g_2^2 = \frac{4}{3} g_1^2.
\eas
The $\w{ij}$ then read
\bas
	\w{11} &= \frac{1}{2},&
	\w{12} &= \frac{9}{20},&
	\w{13} &= \frac{1}{4}, & 
	\w{23} &= \frac{1}{5}.
\eas

\end{rmk}

\section{Supersymmetry of the action} \label{sec:ncmssm-checks}

Even though the three obstructions mentioned at the beginning of Section \ref{sec:NCSSM} are avoided and the particle content of this theory coincides with that of the MSSM, we do not know if the action associated to it is in fact supersymmetric. In this section we check this by examining the requirements from the list in \cite[\S 3]{BS13I}. We will not cover all of them here, however.

Before we get to that, we note that each of the fields $\sfer_{ij}$ appears at least once in one of the building blocks of the third type. This can easily be seen by taking all combinations $(i,j)$, $(i,k)$ and $(j,k)$ of the indices $i,j,k$ of each of the building blocks of the third type that we have. Put differently, there is at least one horizontal line between each two `columns' in the Krajewski diagram of Figure~\ref{fig:MSSM_bb3_compact}. This means that for each sfermion field $\sfer_{ij}$ of the MSSM that is defined via the building block \B{ij}, we can meet the demand \eqref{eq:bb2-resultCiij} on the parameters $C_{iij}$, $C_{ijj}$ that supersymmetry sets on them. We do this by setting them to be of the form 
\ba\label{eq:bb3-expressionCs}
	C_{iij} &= \sgnc_{i,j} \sqrt{\frac{r_i}{\w{ij}}}(N_k \yuks{i}{j}\yuk{i}{j})^{1/2}	
\ea
where $r_i$ and $\w{ij}$ were introduced in \eqref{eq:ncmssm-ident-gauge-kin} and \eqref{eq:defw} respectively, and $\yuk{i}{j}$ is the parameter of the building block \B{ijk} that generates $\sfer_{ij}$ (cf.~\cite[\S 2.3]{BS13I}). With the right choice of the signs $\sgnc_{i,j}, \sgnc_{j,i}$ for these parameters, the fermion--sfermion--gaugino interactions that come from the building blocks of the second type coincide with those of the MSSM.
\begin{itemize}
\item 
For each of the four building blocks \B{11_R2_L}, \B{1_R2_L3}, \B{1\bar1_R2_L} and \B{\bar1_R2_L3} of the third type that we have, there is the necessary requirement \eqref{eq:improvedUpsilons} for supersymmetry. In the parametrization \eqref{eq:bb3-expressionCs} of the $C_{iij}$ these relations read:
\ba
	\sgnc_{i,j} \sqrt{\w{ij}}\,\yukw{i}{j} &= - \sgnc_{i,k}\sqrt{\w{ik}}\,\yukw{i}{k}, &
	\sgnc_{j,i} \sqrt{\w{ij}}\,\yukw{i}{j} &= - \sgnc_{j,k}\sqrt{\w{jk}}\,\yukw{j}{k}, \nn\\
	\sgnc_{k,i} \sqrt{\w{ik}}\,\yukw{i}{k} &= - \sgnc_{k,j}\sqrt{\w{jk}}\,\yukw{j}{k}, &&
\label{eq:improvedUpsilons_ncmssm}
\ea
where we have written
\bas
	\yukw{i}{j} &:= \yuk{i}{j}(N_k\tr\yuks{i}{j}\yuk{i}{j})^{-1/2}, & \yukw{i}{k} &:= (N_j\yuk{i}{k}\yuks{i}{k})^{-1/2}\yuk{i}{k},\nn\\
	\yukw{j}{k} &:= \yuk{j}{k}(N_i\yuks{j}{k}\yuk{j}{k})^{-1/2}. &&
\eas
for the `scaled' versions of the parameters \yuk{i}{j}, \yuk{i}{k} and \yuk{j}{k} of the building block \B{ijk}. Here it is $\sfer_{ij}$ that is assumed to have $R = 1$ and consequently no family structure. (See \cite{BS13I}, Remark 28 for the case that it is $\sfer_{ik}$ or $\sfer_{jk}$ instead.) To connect with the notation of the noncommutative Standard Model, we will write 
\bas 
 \yuk{\nu}{} &:= \yuk{1_R,1}{2_L},& \yuk{u}{} &:= \yuk{1_R,3}{2_L}
\eas
for the parameters of the building blocks \B{1_R12_L} and \B{1_R32_L} that generate the up-type Higgs fields and
\bas
 \yuk{e}{} &:= \yuk{\bar1_R,1}{2_L}, & \yuk{d}{} &:= \yuk{\bar 1_R,3}{2_L}
\eas
for those of \B{\bar1_R12_L} and \B{\bar1_R32_L} that generate the down-type Higgs fields. Furthermore, we write
\bas 
	a_u &= \tr_M \big(\yuks{\nu}{}\yuk{\nu}{} + 3 \yuks{u}{}\yuk{u}{}\big), & a_d &= \tr_M\big(\yuks{e}{}\yuk{e}{} + 3 \yuks{d}{}\yuk{d}{}\big)
\eas
for the expressions that we encounter in the kinetic terms of the Higgses:
\bas
	&\n_{1_R2_L}^2\int_M |D_\mu \hu|^2,\qquad \n_{1_R2_L}^2 = \frac{f(0)}{2\pi^2}\frac{1}{\w{12}}a_u\nn\\
\intertext{and}
	&\n_{\bar1_R2_L}^2\int_M |D_\mu \hd|^2,\qquad \n_{\bar1_R2_L}^2 = \frac{f(0)}{2\pi^2}\frac{1}{\w{12}}a_d
\eas
respectively. (Here, the parametrization of \cite[\S 2.3]{BS13I} is used). The factors $3$ above come from the dimension of the representation $\mathbf{3}$ of $M_3(\com)$. Inserting the expressions for the $\yukw{i}{j}$ the above identity reads for the building block \B{1_R12_L}:
\bas
		&- \sqrt{\w{12}}\Big(\yuk{2,1}{1}\yuks{2,1}{1} + \yuk{2,\bar1}{1}\yuks{2,\bar1}{1}\Big)^{-1/2}\yuk{2,1}{1} \nn\\
		&=  \sgnc_{1,2_L}\sgnc_{1,1_R} \sqrt{\w{11}} \,\yuk{1,2}{1}\Big(2 \yuks{1,2}{1}\yuk{1,2}{1}\Big)^{-1/2} 
			=  \sgnc_{2_L,1}\sgnc_{2_L,1_R} \sqrt{\w{12}} \frac{\yuk{\nu}{}{}^t}{\sqrt{a_u}}.
\eas
For \B{\bar1_R12_L}, \B{1_R32_L}, \B{\bar1_R32_L} it reads 
\bas
		&- \sqrt{\w{12}}\Big(\yuk{2,1}{1}\yuks{2,1}{1} + \yuk{2,\bar1}{1}\yuks{2,\bar1}{1}\Big)^{-1/2}\yuk{2,\bar 1}{1} \nn\\
		&=  \sgnc_{1,2_L}\sgnc_{1,\bar1_R} \sqrt{\w{11}}\yuk{\bar 1,2}{1}\Big(2 \yuks{\bar 1,2}{1}\yuk{\bar 1,2}{1}\Big)^{-1/2} 
			=  \sgnc_{2_L,1}\sgnc_{2_L,\bar1_R}\sqrt{\w{12}} \frac{\yuk{e}{}{}^t}{\sqrt{a_d}},\\
		&- \sqrt{\w{23}}\, \yuk{2,1}{3}\Big(\yuks{2,1}{3}\yuk{2,1}{3} + \yuks{2,\bar1}{3}\yuk{2,\bar1}{3}\Big)^{-1/2} \nn\\
		&=  \sgnc_{3,2_L}\sgnc_{3,1_R} \sqrt{\w{13}}\Big(2 \yuk{1,2}{3}\yuks{1,2}{3}\Big)^{-1/2}\yuk{1,2}{3} 
			=  \sgnc_{2_L,3}\sgnc_{2_L,1_R} \sqrt{\w{12}} \frac{\yuk{u}{}}{\sqrt{a_u}},\\
\intertext{and}
		&- \sqrt{\w{23}}\, \yuk{2,\bar 1}{3}\Big(\yuks{2,1}{3}\yuk{2,1}{3} + \yuks{2,\bar1}{3}\yuk{2,\bar1}{3}\Big)^{-1/2} \nn\\
		&=  \sgnc_{3,2_L}\sgnc_{3,\bar1_R}\sqrt{\w{13}}\Big(2 \yuk{\bar 1,2}{3}\yuks{\bar 1,2}{3}\Big)^{-1/2}\yuk{\bar 1,2}{3} 
			=  \sgnc_{2_L,3}\sgnc_{2_L,\bar1_R}\sqrt{\w{12}} \frac{\yuk{d}{}}{\sqrt{a_d}}
\eas
respectively. We have suppressed the subscripts $L$ and $R$ here for notational convenience and used Remark 28 for the identities associated to \B{11_R2_L} and \B{1\bar1_R2_L}, giving rise to the transposes of the matrices \yuk{\nu}{} and \yuk{e}{} above. Not only do these identities help to write some expressions appearing in the action more compactly, it also gives rise to some additional relations between the parameters. Taking the second equality of each of the four groups, multiplying each side with its conjugate and taking the trace, this gives 
%
\begin{subequations}
\ba
	\frac{M}{2}\w{11}a_u &= \w{12}\tr_M \yuks{\nu}{}\yuk{\nu}{}, &
	\frac{M}{2}\w{11}a_d &= \w{12}\tr_M \yuks{e}{}\yuk{e}{},\label{eq:ncmssm-rel1-line1} \\
	\frac{M}{2}\w{13}a_u &= \w{12}\tr_M \yuks{u}{}\yuk{u}{}, &
	\frac{M}{2}\w{13}a_d &= \w{12}\tr_M \yuks{d}{}\yuk{d}{},\label{eq:ncmssm-rel1-line2}
\ea
\end{subequations}
where on the LHS there is a factor $M$ coming from the identity on family-space. Summing the first and three times the third equality (or, equivalently, the second and three times the fourth), we obtain 
\ba\label{eq:MSSM-condition1.1}
	\w{11} +
	3\w{13} &= \frac{2}{M}\w{12}.
\ea
Similarly, we can equate the first and last terms of each of the four groups of equalities, multiply each side with its conjugate and subsequently sum the first two (or last two) of the resulting equations. This gives
\begin{subequations}
\ba
	\id_M &= \frac{\yuk{\nu}{}^t(\yuks{\nu}{})^t}{a_u} + \frac{\yuk{e}{}^t(\yuks{e}{})^t}{a_d} \label{eq:ncmssm-rel2-line1}\\
\intertext{and} 
	\frac{\w{23}}{\w{12}}\id_M &= \frac{\yuks{u}{}\yuk{u}{}}{a_u} + \frac{\yuks{d}{}\yuk{d}{}}{a_d}\label{eq:ncmssm-rel2-line2}
\ea
\end{subequations}
respectively. By adding the first relation to three times the second relation and taking the trace on both sides, we get
\ba\label{eq:MSSM-condition1.2}
	\w{12} = \frac{3M}{2-M}\w{23}.
\ea	
We combine both results in the following way. We add the relations of \eqref{eq:ncmssm-rel1-line1} and insert \eqref{eq:ncmssm-rel2-line1} to obtain
\bas
	\frac{M}{2}\w{11} +	\frac{M}{2}\w{11} &= \w{12}\bigg(\tr_M \frac{\yuks{\nu}{}\yuk{\nu}{}}{a_u} + \tr_M \frac{\yuks{e}{}\yuk{e}{}}{a_d}\bigg) = \w{12}M,
\eas
i.e.~
\ba\label{eq:ncmssm-condition5}
	\w{11} &= \w{12}. 
\ea
Similarly, we add the relations of \eqref{eq:ncmssm-rel1-line2}, insert \eqref{eq:ncmssm-rel2-line2} and get
\ba\label{eq:ncmssm-condition6}
	\w{13}M &= \w{12}\tr_M \bigg(\frac{\w{23}}{\w{12}}\id_M\bigg), && \text{or} & \w{13} &= \w{23}.
\ea

\item We have four combinations of two building blocks \B{ijk} and \B{ijl} of the third type that share two of their indices \cite[2.3.1]{BS13I}. Together, these give two extra conditions from the demand for supersymmetry, i.e.~that $\w{ij}$ (as defined in \eqref{eq:defw}) must equal $\tfrac{1}{2}$ \cite[\S 3]{BS13I}:
	\begin{subequations}\label{eq:ncmssm-condition4}
		\ba
			\B{1_R2_L1}\ \&\ \B{1_R2_L3}: &\quad \w{12} = \frac{1}{2},\\
			\B{32_L1_R}\ \&\ \B{32_L\bar1_R}: &\quad \w{23} = \frac{1}{2}.
		\ea
	\end{subequations}
		The other two combinations, \B{\bar1_R2_L1}\ \&\ \B{\bar1_R2_L3} and \B{12_L1_R}\ \&\ \B{12_L\bar1_R}, both give the first condition again.

%

\end{itemize}

Combining the conditions \eqref{eq:MSSM-condition1.1}, \eqref{eq:MSSM-condition1.2} and \eqref{eq:ncmssm-condition4} we at least need that
\bas
	\w{11} &= \w{12} = \w{13} = \w{23} = \frac{1}{2}
\eas
for supersymmetry. However, if we combine this result with \eqref{eq:MSSM-condition1.1} and \eqref{eq:MSSM-condition1.2} it requires
\ba\label{eq:ncmssm-doesnotwork}
	2 - M &= 3M &&\text{and}& 4 &= \frac{2}{M}\quad \Longrightarrow\quad M = \frac{1}{2}.
\ea

%

We draw the following conclusion:
\begin{theorem}
	There is no number of particle generations for which the action \eqref{eq:totalaction} associated to the almost-commutative geometry determined by \eqref{eq:ncmssm-acg}, which corresponds to the particle-content and superpotential of the MSSM, is supersymmetric.
\end{theorem}

Since the extension \eqref{eq:ncmssm-acg-ext} of the finite spectral triple with extra Higgs/higgsino copies does not have an effect on which building blocks of the third type can be defined, the calculations presented in this section and hence also the conclusion above are unaffected by this.


%

Does this mean that all is lost? Suppose we focus on further extensions of the MSSM, such as that of Theorem 10 of \cite{BS12}. Since such extensions have extra representations in $\H_F$, this also creates the possibility of additional components for $D_F$. Which components these are exactly, depends on the particular values of the gradings $\gamma_F$ and $R$ on the representations. However, for the extension mentioned above in particular, we can check that for all combinations of values, the permitted components can never all be combined into building blocks of the third type, thus obstructing supersymmetry.

In general, any other extension might allow for extra building blocks of the third type, making the results \eqref{eq:MSSM-condition1.1} and \eqref{eq:MSSM-condition1.2} subject to change. The demands \eqref{eq:ncmssm-condition4} that follow from adjacent building blocks of the third type remain, however. If we add a building block of the fourth type for the right-handed neutrino, this requires $r_1 = \tfrac{1}{4}$ (see Proposition 32 of \cite{BS13I}). This can only hold simultaneously with \eqref{eq:ncmssm-condition4} if
\bas
	r_1 &= \frac{1}{4}, & r_2 &= \frac{1}{8}, & r_3 &= \frac{1}{12}.
\eas
Enticingly, for $M \leq 3$ these required values are all smaller than or equal to the actual ones of \eqref{eq:ncmssm-rs} and \eqref{eq:ncmssm-rs2}, implying that there might indeed be extensions of $\H_F$ for which they coincide.

\section*{Acknowledgements}
One of the authors would like to thank the Dutch Foundation for Fundamental Research on Matter (FOM) for funding this work.

\bibliographystyle{plain}
\providecommand{\noopsort}[1]{}

\end{document}